\documentclass{article}
\usepackage[english]{babel}
\usepackage{epsfig}
\input epsf.tex
\usepackage{graphicx}
\usepackage{amsmath}

\begin{document}

\newcommand{\nqw}{\chi^{(3)}_{\mathrm{QW}}}
\newcommand{\nhyb}{\chi^{(3)}_{\mathrm{Hyb}}}
\newcommand{\lT}{l_{T}}
\newcommand{\vE}{\mathbf{E}}
\newcommand{\done}{d_{1}}
\newcommand{\dtwo}{d_{2}}
\newcommand{\vrr}{\mathbf{r}}
\newcommand{\ror}{\rho_{m}(\vrr)}
\newcommand{\vP}{\mathbf{P}(\vrr)}
\newcommand{\vd}{\mathbf{d}_{vc}}
\newcommand{\vdp}{\mathbf{d}_{vc}^{\prime}}
\newcommand{\vp}{\mathbf{p}}
\newcommand{\ulam}{u_{\lambda}(\vrr)}
\newcommand{\uetac}{u_{\eta}^{*}(\vrr)}
\newcommand{\Icell}{\int_{\Omega} d\vrr \,}

\title{Resonant Energy Transfer from Organics to Quantum Dots
and Carrier Multiplication}
\author{Vladimir M. Agranovich $^{1,2}$\footnote{Corresponding Author: Vladimir Agranovich,
UTD NanoTech Institute, 
P.O. Box 830688, BE26,
Richardson, TX 75083-0688, USA,
 Tel: 972-883-6545,
Fax: 972-883-6529, EMail: vladimir.agranovich@utdallas.edu.}
 and  Gerard Czajkowski$^{3,4}$ \\
$^{1}$\emph{Chemistry Department, The University of Texas at
Dallas, Richardson, Texas 75083, USA}\\
$^{2}$\emph{Institute of Spectroscopy, Russian Academy of Science, Moscow, Russia}\\
$^{3}$\emph{University of Technology and Life Sciences, Bydgoszcz,
Poland}\\
$^{4}$\emph{Scuola Normale Superiore,  Pisa, Italy }} \maketitle
\begin {abstract}

It was shown in the recent experiments that the hybrid
organic/inorganic resonant structures can provide a flexible
materials platform aimed at the design of novel light emitting
devices. The applications of hybrid structures for photovoltaic
solar cell can also be useful. We pay attention in this note
that the resonant energy transfer in hybrid structure from
the organic thin layer to the semiconductor nanostructures can
drastically increase the intensity of the free carrier generation.
To demonstrate this idea we use the results of recently published
paper by Zhang {\it et al.}~\cite{bul} demonstrating the highly efficient
resonance energy transfer from J-aggregates layer to semiconductor
nanocrystals. It is known that the semiconductor nanocrystals with
small energy gap represent a promising route to increased solar
conversion in single--junction photovoltaic cells. We argue that
the using of nanocrystals with small energy gap in the hybrid
organic/inorganic structures similar to created in \cite{bul} can
increase tens times the total intensity of carrier
multiplication. The organic part in such  hybrid structures will
play a role of the peculiar organic concentrator of the light
energy.
\end {abstract}

\section{Efficient Energy Transfer in Resonant\\ Organic/Inorganic
Nanostructures}

Majority of commercial optoelectronic devices are built on the
basis of inorganic or organic semiconductors, respectively, in
light emitting devices, solar cells and nonlinear media.
Crystalline inorganic semiconductors exhibit a good electronic
transport with moderate optical cross--sections. Organic materials
possess large optical oscillator strengths but are mostly limited
to relatively low performance areas due to inferior mobility of
charge carriers. A fundamentally new direction to overcome some of
the limitations and liabilities of conventional material
approaches while creating an entirely different class of
multifunctional high performance optoelectronic materials is to
exploit physical synergy in \textit{resonant} organic-inorganic
hybrid nanostructures. The most common form of exciton in an
organic semiconductor is a Frenkel exciton, which is tightly
localized on a single molecular site and moves as a unit through
the molecules. On the other hand, the Wannier–-Mott excitons found
in inorganic semiconductors can span across hundreds of lattice
unit cells. In resonant organic--inorganic structures the
electronic excitations of the organic part (Frenkel excitons,
strong excitonic transitions) and of the inorganic part
(Wannier--Mott excitons, good transport, large nonlinearities) are
energetically matched and coupled in a weak or strong coupling
regime. The resonant nanostructures could thereby provide a
flexible materials platform aimed at the design of novel
optoelectronic devices exploiting specific assets of both their
organic and inorganic constituents and exhibiting qualitatively
new properties that would be absent in each class of the materials
separately. The different properties of hybrid nanostructures,
including properties of hybrid nanostructures in a weak and strong
coupling regime and also in microcavity configuration, was
discussed in many papers (see, for instance, review papers
\cite{ABLB,A} and references therein). In this note we discuss
only out-of- microcavity hybrid nanostructures in weak coupling
regime. For such structures is typical the possibility of fast and
efficient non-radiative (F\"{o}rster--like) excitation
\textit{energy transfer} between inorganic and organic
nanostructures due to the pure electrostatic interaction.
\begin{figure}[phb!]
\begin{center}
\includegraphics[width=0.9\textwidth]{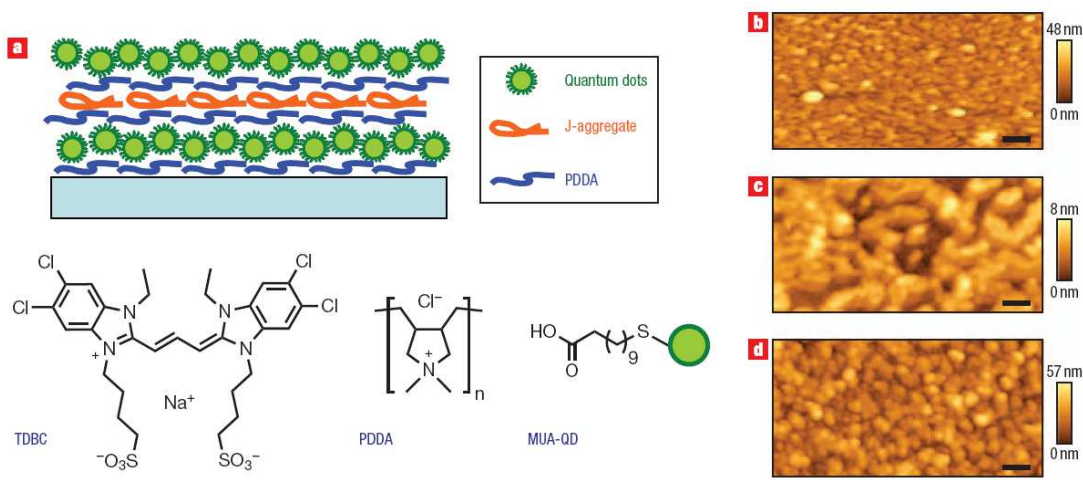}
\caption[Hybrid organic - inorganic multilayer film] {\small
Hybrid organic-inorganic (J-aggregate/QD) multilayer film
deposited by LBL assembly. a, Schematic of the hybrid film layer
structure: a monolayer J-aggregate of 5,6-dichloro-2- [3- [5,6 -
dichloro - 1 - ethyl - 3-(3-sulphopropyl) -
2(3H)-benzimidazolidene] -1-propenyl] -1-ethyl-3-(3-sulphopropyl)
benzimidazolium hydroxide (TDBC) is sandwiched between two
monolayers of CdSe-ZnS QDs, joined by monolayers of
poly(diallyldimethylammonium chloride) (PDDA). Molecular
structures of TDBC, PDDA and MUA are illustrated in the lower
panel. b-d, AFM images of the LBL-grown films, consisting of a
monolayer of QDs ($\lambda_{\rm em}= 653~\hbox{nm}$) (PDDA/QD)
(b), a monolayer of J-aggregate (PDDA/TDBC/PDDA) (c), and the
hybrid film II (d). The scale bars in b-d are 100 nm, from
\cite{bul}.}\label{fig13671}
\end{center}
\end{figure}

This effect was suggested \cite{ABLB1} as a new concept for light
emitting devices where the \textit{electric pumping} of
excitations in the inorganic semiconductor quantum well (QW) would
efficiently turn on the \textit{organic} material
\textit{luminescence} for enhanced light emission from the
material composite. Due to its Coulomb mechanism, the energy
transfer is longer--ranged than direct charge transfer and only
has a requirement of a reasonable exciton energy overlap between
organic and inorganic semiconductor materials to create nanoscale
interfaces with useful electromagnetic properties. Very recent
experiments confirmed the possibility of this type of energy
transfer between nanostructures. We have in mind, for example, the
observation of efficient energy transfer between an inorganic
(InGaN) QW and an adjacent layer of quantum dots (QDs)
\cite{achemann}, between a (Ga, In)N/GaN QW to thin organic
polyfluorene films \cite{itskos}, between ZnO QW to an organics
$\alpha-sexithiophene$ (6T) or POPOP in \cite{blum}. In mentioned
papers the observed coupling takes place between excitons within
the hybrid composite. To fully exploit the non-radiative,
electrostatic coupling between these different types of excitons,
one must seek for materials with large exciton transition dipole
moment and consequently with a large oscillator strength. These
materials, as known, in the same time should exhibit a strong
photon--matter interactions (strong absorption  and short
radiative life time of excitons). These properties help in the
search of materials which can be used in constructing the resonant
organic/inorganic nanostructures.

\section{F\"{o}rster-like Energy Transfer from Organics to
Quantum Dots for Carrier Multiplication}

 The possibility of energy transfer from organics to quantum well
 was noted in consideration of interaction of quantum well excitations
 with a resonant localized excitation in organics \cite{2005}.
 Such process is opposite to
energy transfer from quantum well (or quantum dot) to organic
material which we considered above in the previous section. As we
mentioned, such process can be used for the creation of a new type
of the light emitting devices. The opposite process of electronic
energy transfer from organic material to semiconductor
nanostructure (quantum well or quantum dots) which we consider in
this section can be interesting for a creation of a new type of
solar cell. The same as under influence of photons this process
also will result in creation of a free electron--hole pairs in
inorganic nanostructure which in contrast to organic materials
typically have a high mobility of carriers. Thus, in this case,
the most complicated process for organic solar cells, the process
of the transformation of the Frenkel exciton into the
charge-–transfer exciton with the subsequent charge separation
does not appear at all. However, in many cases the energy pumped
by organic nanostructure to inorganic one can be much larger that
those pumped by photons from the same beam. We demonstrate this
possibility considering as the model of hybrid structure the
results of the recently published paper \cite{bul}. Using this
paper we discuss the electronic energy transfer from organics to
semiconductor nanocrystals (NCs)and its possible role in the
process of charge multiplication. However, in contrast to NCS
taken in \cite{bul} we propose to use NCs with small energy gap.

In the paper \cite{bul} the multilayer structure presented in
Fig.1 has been investigated. The structure was obtained by
layer–-by-–layer (LBL) assembly approach. Figure 1.a schematically
shows the typical structure of the hybrid organic/inorganic
(J-aggregate/QD) LBL films that was synthesized for the discussed
study, along with the chemical structures of the constituents. In
the hybrid film a single monolayer J-aggregate of cyanine dye
(TDBC) was sandwiched between two monolayers of CdSe–-ZnS
core–shell structured NCs (quantum dots), with polyelectrolyte
(PDDA) acting as the ultrathin "molecular glue". Two types of
hybrid films were considered in \cite{bul}, one with NCs emission
centered at 548 nm (referred to as film I) and the other at 653 nm
(referred to as film II), thus providing contrasting cases of
excitation coupling with respect to the fixed J-aggregate emission
at 594 nm. Representative atomic force microscopy (AFM) images of
hybrid film II and two additional NCs and J-aggregate reference
films are shown in Fig. 1. b-d.

It was found for films II that resonance coupling of electronic
excitations in J-aggregates with electronic excitations in the
two monolayers of NCs can reach efficiencies of energy transfer
from J-aggregates to NCs up to 98 percents at room temperature.
This result can be especially interesting for properties of
single-junction photovoltaic cells if in the structure similar
to that presented in Fig. 1. instead of layers of CdSe-ZnS
core–shell structured NCs used in \cite{bul} to use NCs with small
energy gap (PbSe or similar). Such NCs usually are used in the
studies of the carrier multiplication, the effect which is the
direct photogeneration of multiexcitons in NCs by single photons
\cite{nozik,klim,nozik1}.

NCs contain approximately 100 to 10,000
atoms. Because of the strong spatial confinement of electronic
wave functions and reduced electronic screening, the effects of
carrier--carrier Coulomb interactions are greatly enhanced in NCs
compared with those in bulk materials. These interactions open a
highly efficient decay channel via Auger recombination and just a
strong carrier-–carrier interaction in NCs is responsible for
carrier multiplication~\cite{nozik}. It is clear that the F\"{o}rster
resonant energy transfer (FRET) of electronic energy excitation
with energy 2 - 4 eV from organic material to NCs with small
energy gap of order of 0.5 eV in structures similar to presented
in Fig.~1 also can give a carrier multiplication. However, this
process can have a few interesting peculiarities in comparison
with carrier multiplication under influence of photons of the same
energy.

First of all a high efficiency of energy transfer from
organic material to NCs can drastically increase the number of
carriers in NCs, because the absorbtion of light by rather dense organic
material can be much larger than the absorption of light by the
system of separated NCs (in the case of structure created in
\cite{bul} the absorption in organics is larger than the
absorption of NCs system more than  ten times, see Fig. 2 from \cite{bul}).
One can expect that absorption of the NCs based on an equivalent bulk
volume is very close to absorption of corresponding bulk
semiconductor. However, the NCs have to be separated and the relatively
small absorption of light by NCs is the result of a small density of
NCs which cannot be larger than some critical value.

The second important feature of the carrier multiplication under
the influence of energy transfer is the structure of electric
field in the volume of NC. In the simplest approximation this
field is created by transition dipole moment in organic molecules.
This field even in spherical NCs is more inhomogeneous and
non–-spherical than created by plane wave of photon (see, for
example, \cite{BornWolf}, Ch. XIII). It is clear that this
asymmetry of the electric field could be responsible for change of
the selection rules, determining the population of electron higher
energy states in NC. This effect needs a careful analysis and is
interesting because it can decrease the minimal energy of
excitation which produces carrier multiplication. In the case, for
example, of pumping of NCs (PbSe and PbS) by light this minimal
energy is equal to the threefold energy gap \cite{klim, nozik1}.
One can expect that for such NCs and with pumping of them by
F\"{o}rster--like energy transfer this threshold will be smaller.
\section*{Acknowledments}
V. M. A.  thanks Anton Malko and Yuri Gartstein
for discussions. He is also thankful to Russian Foundation of Basic Research
for partial support.

\end{document}